\documentclass[aps,prl,amsmath,amssymb,sort&compress,comma,super,nofootinbib,floatfix]{revtex4-1}
\usepackage{graphicx}
\usepackage{epstopdf}
\usepackage{subcaption}  

\begin{document}

\newcommand{\etal}{{\it et al.}\/}
\newcommand{\gtwid}{\mathrel{\raise.3ex\hbox{$>$\kern-.75em\lower1ex\hbox{$\sim$}}}}
\newcommand{\ltwid}{\mathrel{\raise.3ex\hbox{$<$\kern-.75em\lower1ex\hbox{$\sim$}}}}
\renewcommand{\thefootnote}{\alph{footnote}}

\title{A Walk-Through of AGN Country --- for the somewhat initiated!}
\author{Robert Antonucci}
\affiliation{Physics Department, University of California, Santa Barbara, CA 93106-9530, USA}

\date{\today}

\begin{abstract}

Some key issues in AGN and galaxy formation are discussed. Very
successful Unified Models explain much of the variety of AGN with orientation
effects; ingredients are shadowing by a dusty ``torus" and relativistic beaming.
A spinoff result is described which is important for the formation of massive
elliptical galaxies. It's the most spectacular and unequivocal AGN feedback
phenomenon known. This is the so-called ``alignment effect" in powerful radio
galaxies at $z\gtwid1$. One of them is a BAL radio {\it galaxy}! I explain a
very robust derivation of the reddening law for nuclear dust, which reveals a
dearth of small grains. Then the quasistatic thin accretion disk model, thought
by many to explain the energetically dominant optical/UV continuum, is thoroughly
debunked. Much of this was known when the model was proposed 35 years ago. A new
argument is given that trivially falsifies a huge superset of such models. It's
possible to see the central engine spectrum with the atomic and dust emission
surgically removed! Few noticed this breakthrough work. The far IR dust emission
in Cygnus A is 10\% polarized, and so far high nuclear dust polarization is seen
in all radio loud objects, but no radio quiet ones.

\end{abstract}


\maketitle


\section{Introduction}\label{sec:1}

Introduction to the content:

The first two parts of this essay derive from a meeting presentation, my task
being to tell the origin story for the AGN spectroscopic unified model, and
comment on various aspects of its development. There is a second aspect of
geometrical unification of accreting supermassive black holes, by relativistic
beaming. Beaming unification was a compelling and necessary idea, but plagued
with statistical anomalies until Peter Barthel found the solution by ``cheating"!
He read outside his field! He read our optical polarization papers, and
suddenly underwent a religious conversion and started the Unification Church.
I joined the Church right away, and we preached that spectroscopic unification
married Beaming Unification, and the two live happily together today. 

I recount some of the sad and almost unbelievable history of modeling
the energetically dominant Big Blue Bump continuum component in Quasars and
other Type 1 AGN. The story is sad because vast theoretical effort has gone
into modeling this Big Blue Bump continuum with optically thick, geometrically
thin and quasistatic accretion disks, despite its gross inconsistency with the
data. Cognitive dissonance was part of the model since the very beginning in the
1980s: it was pointed out explicitly that the variability and polarization were
comically at variance with predictions. Over the 40 years (and counting) of
its ``development," additional profound discrepancies have been discovered, not
least of which is measurement of the surface brightness of the emitter with
microlensing: it is an order of magnitude less than the value for opaque
material that can fit the spectral energy distribution.    

References for the variability, polarization, and surface brightness are
Alloin \etal\ 1985 \cite{Alloin:1985}; Antonucci 1988 \cite{Antonucci:1988},
Antonucci, Kinney and Ford 1989 \cite{Antonucci:1989}; and
Dexter and Agol 2011 \cite{Dexter:2011}.

Carl Sagan (1987) wrote, in The Burdens of Skepticism:

{\em
``In science it often happens that scientists say, `You know, that's a really
good argument; my position is mistaken,' and then they actually change their
minds and you never hear that old view from them again.  They really do it."

``It doesn't happen as often as it should, because scientists are human and
change is sometimes painful.  But it happens every day.  I cannot recall the
last time that happened in politics or religion."
}
{\sc [The Skeptical Inquirer 12, 1, 1987]}

My experience has been that when I argue individually for the rough validity
of the Unified Models, my audience finds me persuasive and tends to adopt it
as a very helpful approximation, and to acknowledge its prodigious predictive
power.

By contrast, I've had a great many opportunities in papers and talks to detail
the profound falsification of the ``standard" disk model.  In that case, my
listeners appear to be similarly moved by the force of the arguments, but revert
their views within at most about three days.
They relax back into Magical Thinking.  I've measured that time constant at
meetings. Two or three days and the Ostrichs' heads are back in the sand.
They resume the nearly universal practice of referring to ``the accretion disk"
and the Big Blue Bump (BBB) optical/UV continuum component synonymously. I now
stop reading when I get to such a statement in a paper. It means that the
authors are unaware or uninterested in the observations.

I've made clear since my first critiques in the 1980s that accretion by black
holes is almost certain to be the energy source for AGN; and that I'd bet my
car that there is a flat isodensity contour somewhere in the inner accretion
flow. But alas we are not looking at the photosphere of a disk which remotely
resembles the standard version that almost everyone accepts, along with the scaling
relations that it implies.

\vspace{5mm}\hrule\vspace{5mm}\par

Instead of forcing my content to fit a structure, I will just take you on a walk
through AGN country, and tell you about some things that interest me.
In keeping with the intent of this volume, I will remind you of a precious
opportunity to separate radiative from dynamical effects of an accreting
massive black hole on its host. I will also draw attention to a spectacular
body of knowledge about giant Elliptical galaxy star-formation in powerful
radio galaxies at redshifts $1\ltwid z\ltwid 4$, which illustrates feedback in
the clearest possible way, and at the greatest possible scale. This is the radio/optical
``alignment effect". (Tim Heckman
tells me galaxy formation people pay no attention to this incredible
phenomenology, because they labor under the illusion that the extended
optical light is just scattered from hidden quasars! ) I'll give seemingly
powerful testable arguments that something akin to a latter-day monolithic
collapse event (or cooling flood) occurs in certain cases. I will highlight a
literal Broad Absorption Line (BAL) radio GALAXY, not a quasar! --- surely the
most spectacular Galactic Wind ever seen. Yet these amazing aspects of AGN
feedback are known to very few people in that field, presumably due to the
usual balkanization by sub-speciality that holds back all fields of science.  

Very fundamentally, the standard disk model predicts a proportionality between the
characteristic temperature and the product $L^{1/4} M^{-1/2}$. This is required of {\it any} model with
thermal radiation from a fixed region set by ${\rm Rg}^2$ (that is, ${\rm M_{bh}}^2$).
This has been utterly falsified multiple times. In fact in the present paper, I explain how the reader can
exclude such models with a figure perfect for a T-shirt: A mighty river of
theory papers flowing for 40 years, negated by a simple observation based on
published literature. An enterprising person can disprove that dependence on
$M$ and $L$\dots or anything else!

I introduce the reader to the five extant quasar spectra freed of the awful
contamination from surrounding reprocessed emission in the various emission
line regions and the dust torus. These spectra don't look like 
the spectra of quasars which you have seen. But they do look {\it very} much
like each other!

I will prove that many AGN are heavily absorbed by dust which is gray in the
UV! You don't have to believe this result, but if not, you have to give up an
Axiom which may be dear to you!

Finally some recent and quite unexpected results are given regarding the
wonderful Cygnus A radio galaxy, and on a very intriguing apparent difference
between radio loud and radio quiet AGN. They tentatively differ systematically
in their magnetic fields on pc scales!

\section{Part 1. How I became Mr.~Doughnut. --- the Spectroscopic Unified Model.}

The AGN Mesoscale is my new name for the region around the dust sublimation
radius through say the Narrow Line Region, where the AGN is still driving the car.

The discovery of the first hidden Type 1 nucleus with polarimetry, the Narrow Line
Radio Galaxy 3C234.    



\subsection{a) ``Take out the dark slide."}

After I published the first hidden Type 1 spectrum in 1984, I met a fellow at a
conference who excitedly and correctly described my paper, saying that they
had discussed it in their journal club and he was sorry he didn't remember the
name of the author. When I confessed I was the author, he said, ``Really?
You're Mister Doughnut?" Well here is how I found the doughnuts with the Type 1
nuclei inside.

When I was a grad student at U C Santa Cruz in the late 1970s, my eventual
advisor, J S Miller, along with postdoc Gary Schmidt, were building an
instrument that could sort photons by polarization as well as wavelength.  At
that time there had been very little spectropolarimetry of AGN. Perhaps it was
considered overkill to get high spectral resolution, since AGN polarization was
known only for identifying broadband processes like synchrotron emission and
scattering.  And so many photons would need to be collected! In fact,
polarimetry was considered by most to be an esoteric fringe field.  

I'd been considering asking Miller to take me on as a graduate student when I
attended a AAS meeting and heard a talk by the director of Kitt Peak National
Observatory, the famous Geoffrey Burbidge. His topic was the wonders that might
be revealed by a truly giant telescope (say 8 m).  On Burbidge's grand agenda
for the futuristic telescope, the last sub-sub-sub-topic was ``Astrometry and
polarimetry." Burbidge may have devoted a dependent clause of his talk to
these two despised specialities, but I think it was really rather a grunt. 

I knew that aside from the well-explored blazar class of synchrotron emitters,
the percent polarizations of AGN are discouragingly low; 1\% is considered to
indicate an attractive target.  And for spectropolarimetry, I knew that to
achieve the same SNR in polarized light as total light, with 1\%
polarization, it would take an exposure longer by a factor of at least 20,000.  

Also I wondered if such low polarization might be due to ``weather," and carry
no important information.
Then in 1979, Stockman \etal\ (1979) \cite{Stockman:1979} reported that the optical
polarization ``vector" in lobe-dominant radio quasars tends very strongly to
align with the $\sim300$ kpc radio axis.  That was electrifying to me because it
meant that the very small optical polarizations were not meaningless weather, but something
fundamental that connected the tiny optical region of perhaps a few a.u. to the
giant radio lobes on scales billions of times larger.  The rough axisymmetry is
maintained from relativistic scales to intergalactic scales!

Joe took me on. I had no talent or interest in instrumentation or software, and
when he showed me the precious incubating instrument, I could only nod and smile
uncomfortably. It looked just like any other gray box of parts to my glazed eyes.
I was no great observer, either. I prepared hugely and relied on incredibly
detailed checklists. For example, before beginning an exposure, I read aloud solemnly,
``Step 1.  Open Dark Slide."
Miller and Schmidt were both very intuitive at the telescope and I later learned
that after that first run, Gary told Joe, ``You've got to get rid of this guy.
Did you see he had to read his checklist to figure out you have to open the
dark slide?"

With linear polarimetry you get 3 of the 4 Stokes' Parameters; instead of just
I, you also measured Q and U simultaneously, with only slight loss of light.
Joe and I, and I guess Gary, had a personality trait in common. We all wanted a
free lunch. A stairway to heaven! We had little assurance that Q and U encoded
a ton of interesting information, but we were all\dots optimistic guys.

\subsection{b) Pay dirt: The Spectroscopic Unified Model, with obscuring torus and polar
periscopic mirror.}

Joe suggested Radio Galaxies for my thesis topic, which I eagerly accepted. We
observed almost every dark run, at the modest-sized 3-m telescope of Lick
Observatory. The site was poor, and the quantum efficiency was also low by
today's standards. When I expressed anxiety about the SNR situation, Joe said, 
``Think of it this way.  We have a big telescope, only it's in series instead of
parallel." Our ``Image Dissector Scanner" was a real Rube Goldberg device,
but devilishly clever in using slightly persistent phosphors on the back of an
image tube, which allowed the single channel detector to zip around, steered
magnetically, and visit all the wavelength channels before they stopped glowing.
The fact is, Lick was eating everyone's lunch with that thing. It was before CCDs. 

During each run we included a few of my radio galaxies, and it was discouraging
work. A series of them showed P consistent with zero, or with just the
polarization due to dichroic absorption in the Milky Way galaxy. The dominant
signal was from stars in the host galaxies. 

One advantage of that crazy instrument was that it truly counted photons, unlike
CCDs, and you could see each one cause an increment in the flux in some
wavelength channel in real time. The spectra for each polarization position angle
were displayed on a small round green oscilloscope screen. Finally we observed 3C234, and we saw the
photons come in, sometimes to the Ordinary Ray spectrum, and sometimes to the
Extraordinary Ray.  Any difference meant polarization was detected.

We sensed a possible detection such that the $0^\circ$ ($=180^\circ$) ray was
stronger than $90^\circ$. I recall that we watched the photons barrel in and
get counted, incrementing the displayed spectra. Ordinary. Ordinary.
Extraordinary. Then we (or at least I) started rooting aloud for each +Q ($0^\circ$) photon,
while cursing those at $90^\circ$.

The Ordinary ray of the Q Stokes Parameter was getting farther ahead! 

Finally Joe said calmly, ``Pay dirt."

Because until that moment, we had {\it nothing}.

\vspace{5mm}\hrule\vspace{5mm}\par

The 3C234 radio galaxy has a jet position angle of $68^\circ$ degrees, to which
I'd assigned an uncertainty of $5^\circ$, which was somewhat arbitrary and a bit
generous. We measured a polarization position angle of $159\pm1^{\circ}$,
almost exactly orthogonal to the radio axis, and we eventually established a
``perpendicular" class of AGN. That is, the optical polarization position
angle in this group was perpendicular to the radio axis. That was opposite to
the discovery of parallel polarization in quasars by Stockman \etal\ And the
magnitude of P was at least ten times higher in the perpendicular group. It became
clear that they were none other than the Type 2 spectral class members.

\begin{figure}[htbp]
\includegraphics[height=8.5cm]{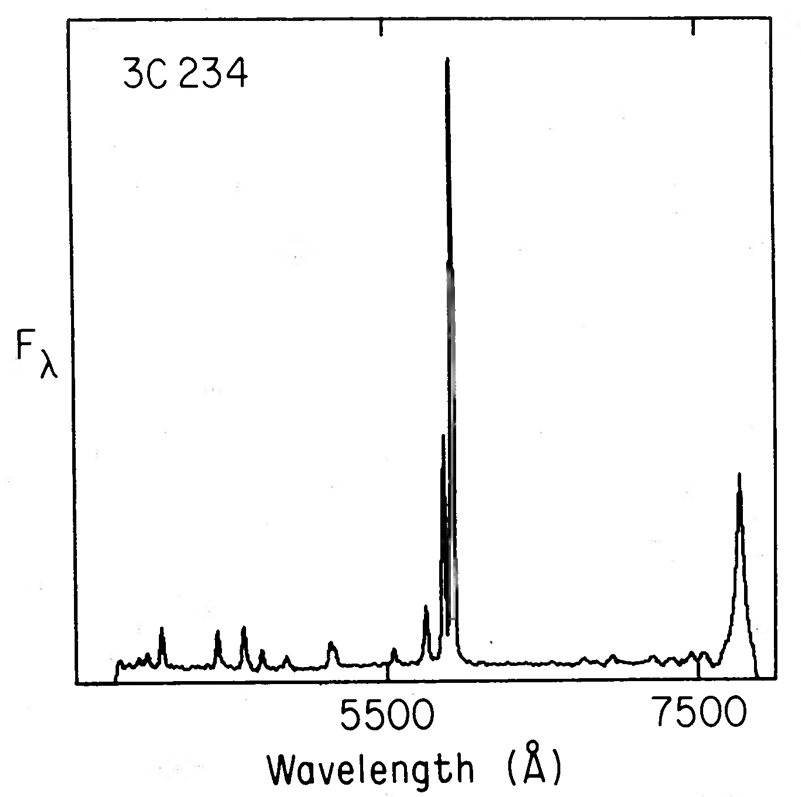}
\caption{Total flux spectrum of the narrow-line radio galaxy 3C234.\label{fig:1}}
\end{figure}

\begin{figure}[htbp]
\includegraphics[height=8.5cm]{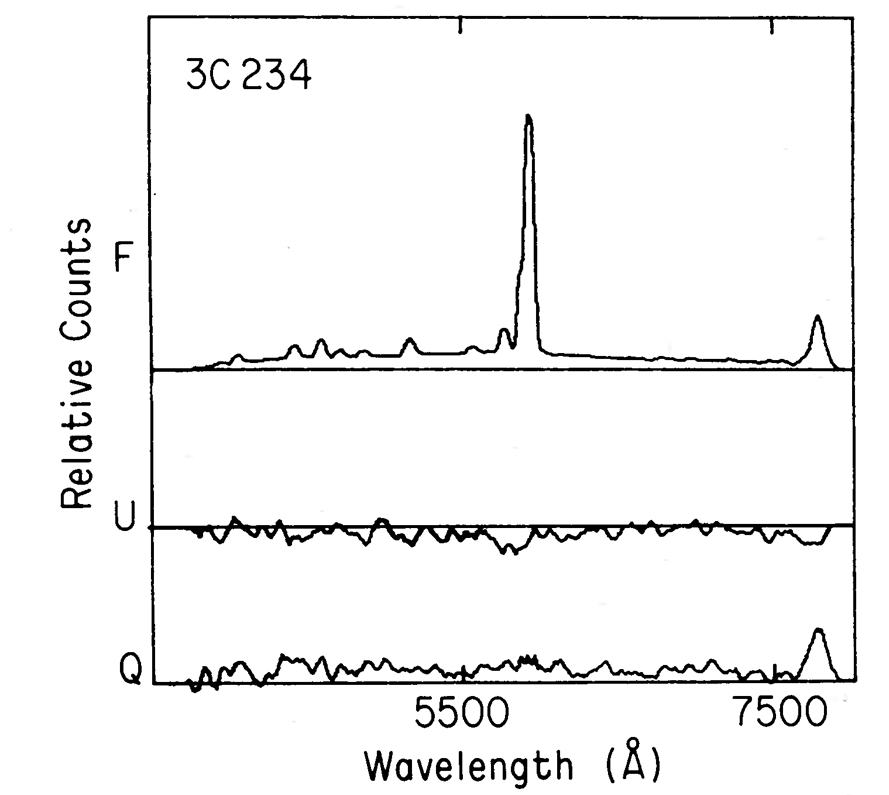}
\caption{The smoothed unnormalized Stokes parameters and count spectrum of 3C234.\label{fig:2}}
\end{figure}

The total flux spectrum of 3C234 is shown in the present Fig.~\ref{fig:1}.
At home, I made
the present Fig.~\ref{fig:2}, which shows the Ordinary minus Extraordinary ray, i.e.
Stokes Q; and then the same for U. As a reminder, this un-normalized Q is the
flux at $0^\circ$ (Ordinary) minus that at $90^\circ$ (Extraordinary) and U is
the same except the $45^\circ$ ray is compared to the $135^\circ$ ray.

The Stokes spectra have been smoothed obviously, but it's very clear that the
[O III] 5007~\AA\ line hardly shows up in the difference spectrum and is thus
unpolarized. But the H$\alpha$ line behaves differently! It is highly polarized, and
accounting for unpolarized narrow components and starlight, the broad component
polarization is a whopping $\sim20$\%.

Emission lines are only intrinsically polarized in very special circumstances,
so I realized this meant we went looking at a hidden broad line (high density;
permitted lines only) region, through a periscopic mirror!

In scattering polarization, physics tells us that {\it the perpendicular to the
position angle is the direction of the photon's last flight before being
scattered into the line of sight}. Thus these photons could have escaped the
hidden nucleus only if they were travelling {\it parallel} to the radio axis. Since
line emission is intrinsically quite isotropic, it follows that the other
directions, the equatorial directions, must be blocked. That is, an opaque
equatorial structure must be hiding the Type 1 nucleus. A correlary is that
astronomers in those regions of the universe with polar views of 3C234 believe
it is a Type 1 object, aka a Broad Line Radio Galaxy.

If this were to be generic, it would explain how radio galaxies could produce
radio lobes without a visible central engine. Both spectral types do have the
engine, and which type of object you saw was an accident.\footnote{You will see
below that this is the whole story for very luminous radio sources,
but many weaker radio galaxies don't participate in this scheme. Very detailed
information can by found in a major review, Antonucci 2012 \protect\cite{Antonucci:2012}.}

I may not know instrumentation but I know about data and systematic errors.
Our setup symmetrized everything in time with respect to light paths,
modulating at about 1~Hz, and the data from that instrument were clean.
Furthermore we were meticulous about correcting for dichroic interstellar
polarization contamination.

I gathered some clues about the wide applicability of the 3C234 scenario.
While the spectropolarimetry paper didn't come out until 1984, I'd published a
short paper in Nature asserting the existence of a class of radio galaxies with
a high perpendicular polarization two years earlier in 1982 \cite{Antonucci:1982}. I also reported
there a parallel class, related to the radio quasars in Stockman \etal\ 1979 \cite{Stockman:1979};
and the following year I did the same for Seyferts. (In the Seyfert case, the
competition missed the parallel and perpendicular polarization alignment groups,
because they combined Types 1 and 2, and took insufficient care with spurious
signals due to interstellar polarization. In fact I only used their published data to
find the alignments. I just analyzed it more carefully.

The geometry described ``unified" the broad line (Type 1) and narrow line
(Type 2) objects, both radio loud and radio quiet. (The language isn't helping
here, because the broad line objects actually have the narrow component as well.)
The term ``unified" is used very specifically: we asserted that if we changed
our viewing angle relative to the AGN axis, the appearance and classification
would change. Thus we may only have one physical class to try to understand.
(The Broad Line or Type 1 radio galaxies can be considered just low luminosity
radio quasars; and now the community recognizes that even powerful quasars are
often observed in the Type 2 orientation.) We didn't use the term obscuring
``torus," but ``thick disk," to avoid any implication regarding the outer
boundary of the structure, or for that matter, of any particular physics at all.
But I've fallen into the common usage of ``torus" here. Keep in mind that the
AGN torus is {\it defined as the structure that blocks photons from the nucleus
which are emitted in the equatorial directions}, based on the radio axes, and
nothing else at all. The torus-like shadowing can be realized by a variety of
physical and kinematical models. The term is used loosely in the literature,
leading to imaginary paradoxes at times.

\subsection{c)  NGC1068, the radio quiet prototype!  why it's good to know nothing.}

Now we turn our attention to radio quiet AGN.  We discovered the nature of the
radio-quiet prototypical Type 2, NGC1068, in two steps.  This was necessary
because certain properties made the interpretation less obvious than for 3C234,
though in retrospect the differences were very superficial. While smarter
astronomers would not have been confused by this\dots alas the task fell to us.
In step 1, published in 1983, we showed that the very strong host contribution
caused the \% polarization to be very low in the optical, with a lot of structure. The
sharp rise towards the UV\footnote{**When I first showed Joe that starlight
heavily dilutes the polarization and strongly affects the wavelength dependence,
he pointed out to me that there were bumps in \% polarization corresponding to the stellar
absorption lines, due to reduced dilution compared to neighboring wavelengths.
He said the \% polarization plot ``looks like a galaxy spectrum plotted upside down."}
simply resulted from dilution of nuclear light by an old stellar population.

Correcting for the effects of the host produced a quasi-power law nuclear
spectrum with constant 16\% polarization. A very similar result was shown very
soon afterwards by McLean \etal\ (1983) \cite{McLean:1983}.
But I was fixated on one tiny bump in \% polarization in the {\it uncorrected} P spectrum,
just redward of Hb, that couldn't be explained by the ``upside down" stellar
absorption lines; nor by dilution by low-polarization narrow emission lines.
It was clearly present in the McLean \etal\ data too, but they didn't mention it,
and they concluded that the whole polarized continuum was due to synchrotron
emission. In our 1983 paper, we mentioned that possibility but hedged, and
favored Thomson scattering. Just as we had a big telescope by using our little
one in series, we did the same with our brains, and published the polarized
flux showing the Seyfert 1 only in 1985.


In almost all cases, huge host contamination makes life harder, but this was an
exception. The excess in the bump in the UN-corrected \% polarization spectrum is due to a
redshifted broad and highly polarized H$\beta$ line component. It {\it only} showed up
in the uncorrected P spectrum, because it reduced the starlight fraction at that
wavelength. After starlight correction, since this scattered broad H$\beta$ has the
same polarization as the nuclear continuum, it's entirely invisible, and indeed
one might assume the p is due to synchrotron radiation. Why was this
``phantom" feature only conspicuous on the {\it red} side of narrow H$\beta$? Because
the scattered line is redshifted by the moving mirror effect. The mirror gas
must be moving away from the nucleus at about 400 km/s, while the narrow line
peaks are blueshifted by $~200$ km/s relative to the systemic redshift.

So 1) look at your data carefully in every form you can think of, and 2) never
get tired of staring at each bump and wiggle till you understand it.

\vspace{5mm}\hrule\vspace{5mm}\par

When one looks through a cosmic mirror with polarimetry, the question arises
as to the nature of the scatterers. The practical possibilities are dust grains
and free electrons. In a sense, the former has greater plausibility because
``dust has orders of magnitude higher cross-section than electrons, per gm of
diffuse matter." Fortunately I wasn't so steeped in this folklore, which is
somewhat dubious anyway. It says that in fully ionized gas with a normal
complement of standard Milky Way grains, dust would greatly dominate the
scattering in the optical/UV region. But that's not a common situation, and in
hot or strongly irradiated media the dust is absent, and the mirror in NGC1068
is like that. Without going into all the arguments, I simply note that
scattering by a mix of grain sizes like that in our Galactic disk scatters
short wavelengths much more efficiently than long wavelengths, and can also
introduce albedo features, especially in the $~2000$~\AA\ region.

\begin{figure}[htbp]
\includegraphics[height=8.5cm]{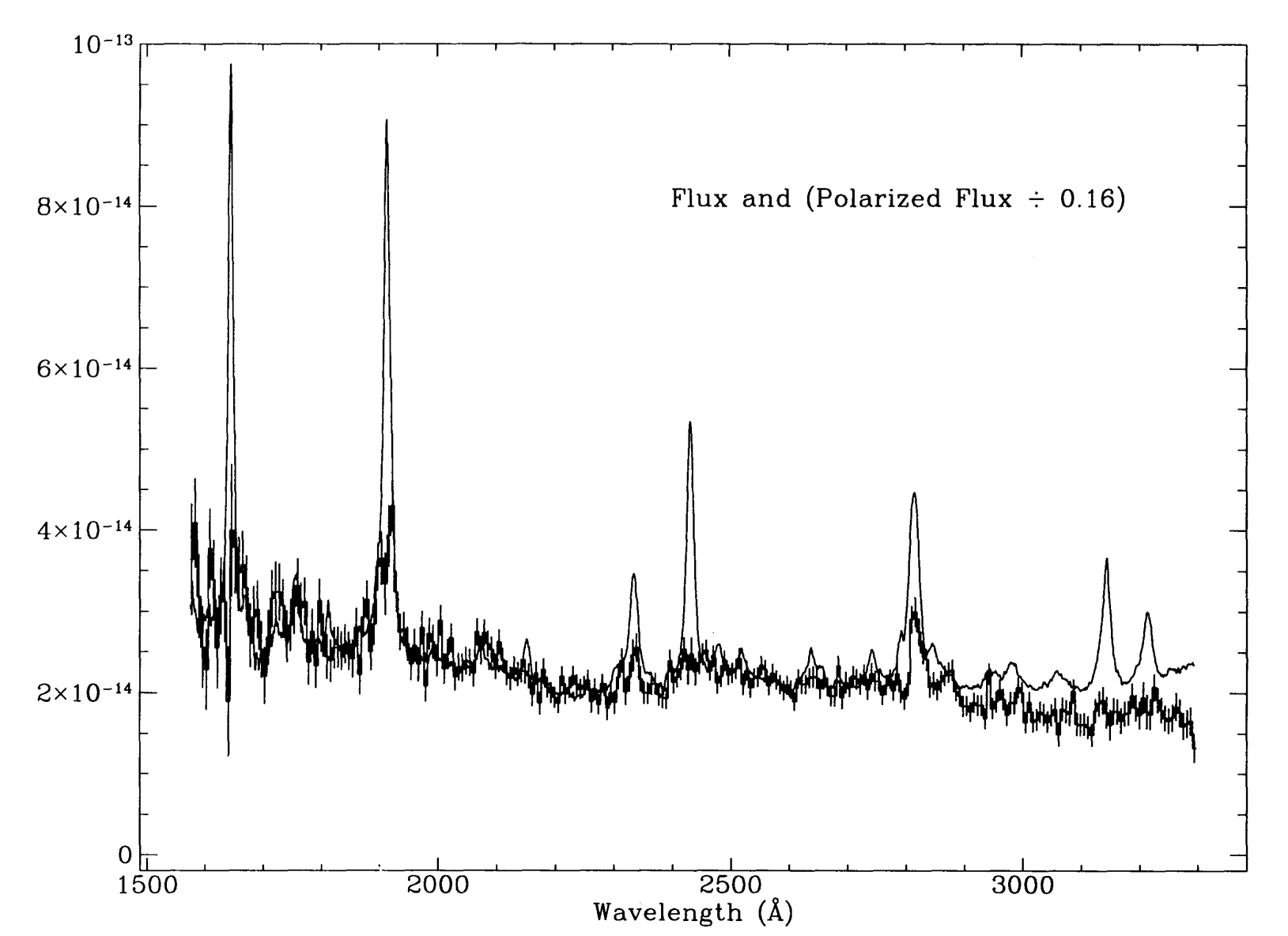}
\caption{NGC1068 ultraviolet polarized flux scaled by 1/0.16 and
plotted over total flux. See text.\label{fig:3}}
\end{figure}

In NGC1068 all the evidence favors Thomson scattering in the central hundred pc,
as we found in Miller and Antonucci (1983) \cite{Miller:1983} and
Antonucci and Miller (1985) \cite{Antonucci:1985}. The
former paper shows that while the {\it observed} \% polarization is low in the
optical in ground-based apertures rising steeply into the UV, careful
correction of the effects of the unpolarized host galaxy leads to a nuclear
continuum with wavelength-independent P at a much higher value, namely 16\%.
We must have done a fantastic job of correcting the optical light for host 
starlight because HST observations in the UV, past the starlight, show exactly
that behavior (Fig.~\ref{fig:3}). Note the overlay of total flux {\it divided by} the
0.16 fractional P from our 1985 paper, on top of the measured UV polarized flux.
The match is perfect, without even any normalization.

That's a signature of electron scattering, and in fact the polarized light looks
exactly like a typical Type 1 with no shape changes. But in science one would like to
prove things at least ten ways, and other strong arguments can be brought to
bear. And as suggested by the ``Proof" of beaming unification below, a direct
argument is better than a consistency check. We suggested electron scattering
for NGC1068, and that raised a flag for another reason. That process normally
dominates the opacity only at very high temperatures, and by ``normally" I mean
in the common case of collisional ionization equilibrium. But we know that the
putative scattering electrons couldn't be very hot, or they would
scatter---broaden any emission lines beyond recognition. We set a very
conservative upper limit on their temperature of a million degrees, but a
hundred thousand would have been justified.  

Krolik and Kallman (1987) \cite{Krolik:1987} then showed that the electron
temperature could be low enough to fit the data, but only if the mirror gas
were highly photoionized. (This is also concluded in
Miller \etal\ (1991) \cite{Miller:1991}, thanks to the wonderful late Bill
Mathews.) And in that case, they said it must produce an enormous $\sim$keV
equivalent-width Fe K$\alpha$ line, whose energy would indicate extremely high
ionization. This was a very dramatic prediction since nothing like that had ever
been seen before. Conversely, they pointed out that if that prediction came true,
then the gas producing the Fe lines would necessarily have significant Thomson
scattering optical depth! (Their predicted line strength was raised further
by Dave Band \etal\ (1990) \cite{Band:1990}, who added the contribution of
resonant scattering.) The prediction was gloriously confirmed by subsequent
X-ray observations. Even better, in a landmark paper by
Ogle \etal\ (2003) \cite{Ogle:2003}, observations and analysis
were presented from the Chandra CCD spectrometer, whose high spatial and spectral resolution
resolved the X-ray narrow line region very well.  Detailed analysis showed that
indeed the emitting clouds {\it must} have about the anticipated product of
Thomson depth and solid angle. Best of all, the radiative recombination
continua directly indicated by their energy width that the electron temperature
was also quite consistent (as low as) that required by the data. 




After I left Santa Cruz, Miller and Bob Goodrich performed some very elegant
and important observations, written up with a big assist by theorist Bill Mathews.
The broad line profile in polarized flux from the nucleus must be a convolution
of the intrinsic one and the electron thermal velocity distribution, and we didn't know
how big an effect the latter factor was. We didn't know whether our mirror
was hot or cold. So Joe and Bob decided to look at the nucleus {\it as
scattered by dust clouds a few hundred pc away}. They knew dust scattering would
reveal the true line profile. They succeeded in measuring it from three different
off-nuclear dust clouds. The line was clearly narrower, as expected!
And the polarized flux was much bluer than from the region where we identified
electron scattering! Their amazing paper \cite{Miller:1991} is called ``Multidirectional Views of
the Active Nucleus of NGC1068.\footnote{*I had the pleasure of showing their\
data at a meeting in 1986. Afterwards Chris McKee said to me, ``I didn't know
telescopes were that powerful!"}

\vspace{5mm}\hrule\vspace{5mm}\par

\begin{figure}[htbp]
\includegraphics[height=8.5cm]{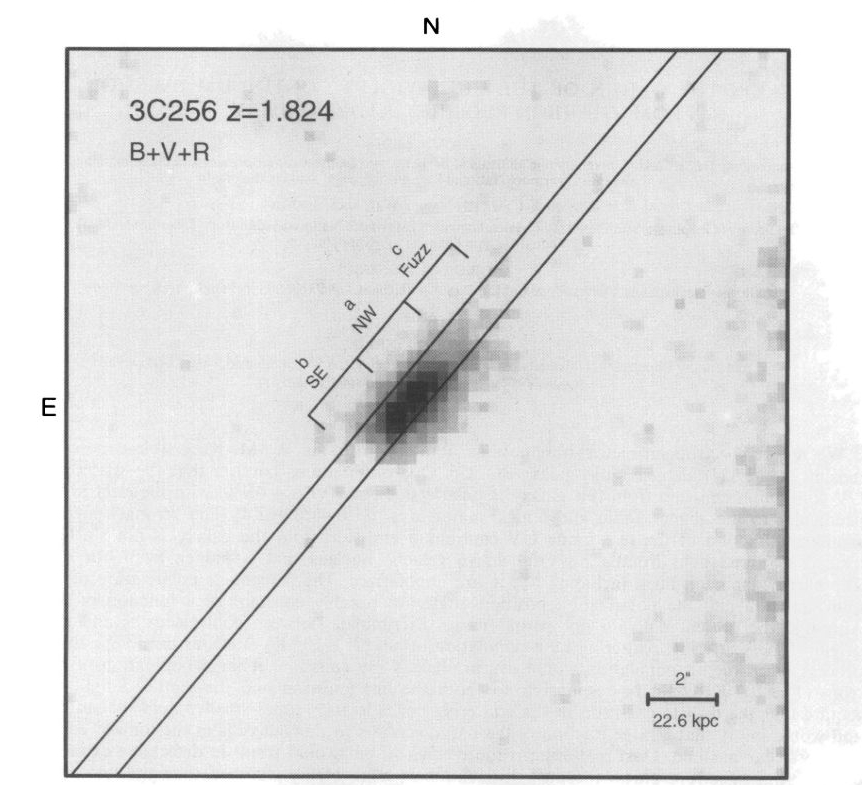}
\caption{Broadband image $(B+V+R)$ of 3C256 obtained by S.~Djorgovski
and H.~Spinrad. See Dey \etal\ (1996) for details.\label{fig:4}}
\end{figure}

\begin{figure}[htbp]
\includegraphics[height=8.5cm]{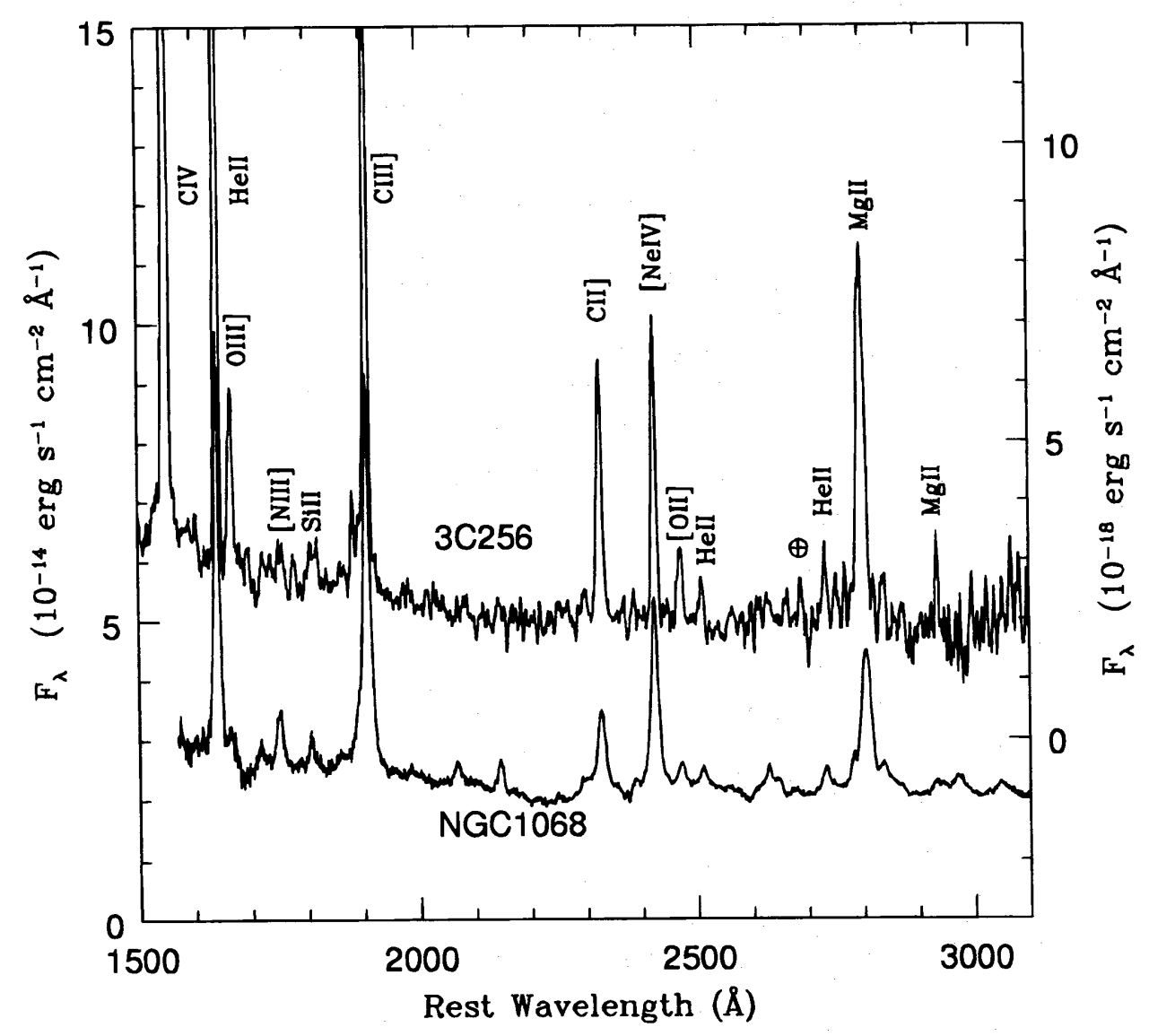}
\caption{Comparison of the rest frame UV continuum spectra of 3C256 and
NGC1068 (from Antonucci, Hurt, \& Miller 1994). The ordinate is
labeled with the flux density scales for NGC1068 and 3C256 on
the left and right axes, respectively. Note that the spectrum of
3C256 is sampled in a $45~{\rm kpc}\times11.3~{\rm kpc}$
aperture ($3''.8\times1''$) whereas the spectrum of NGC1068 is
from a nuclear region of size $471~{\rm pc}\times153~{\rm pc}$
($4''.3\times1''.4$). See Dey \etal\ for additional details.\label{fig:5}}
\end{figure}

The mirror is quite interesting and exotic, but the amazing thing to me is that
some mirrors associated high radio galaxies at relatively high $z$ behave
indistinguishably (see Fig.~\ref{fig:4} for the optical image). Figure~\ref{fig:5}
shows a near-perfect match between C3256 and NGC1068.
In that latter case, the tiny electron scattering cross-section is consequential,
because these mirrors are 10~s of kpc in size, and such a mirror needs to have an
enormous mass, of order $10^{11-12}$ Mo. It's especially interesting because we
argue for 3C256 (Dey \etal\ 1996 \cite{Dey:1996}) that
that enormous mass of baryons is too cool for hydrostatic support. There is no
rotational support either, so one wonders whether all that mass is undergoing
some kind of monolithic collapse! {\it This galaxy differs from the other
``alignment-effect" radio galaxies in lacking a massive old stellar population.}

If I'd been more sophisticated, I'd have discounted electron scattering in both
NGC1068 and 3C256.

\section{Part 2. Unification of the Radio Loud AGN by Relativistic Beaming.}

How the Spectral Unified  Model saved the Beaming Unified Model, and then the
two lived happily ever after. Two ``proofs."

\subsection{a)  Self-calibration, Diffuse Emission, and ``Proof" of Unification
by Relativistic Beaming}

When I was starting out in the 1970s, one of the many attractions of the field
of AGN was Superluminal Motion. In low-frequency radio surveys of the sky,
most of the extragalactic objects found turned out to be giant $\sim 100$ kpc
double-lobed sources. Their steep radio spectra reflect the optically thin
synchrotron emission from the lobes.

But a few were quite different, only showing point sources on the arc-second
resolution maps of the day.  These tiny but very bright little guys showed
flat spectrum (self-absorbed) synchrotron emission, which invited examination
with VLBI networks. Typically this revealed a stationary core and a string of
dots on one side. Re-observation, say a year later in the rest frame, shockingly
revealed that the dots had apparently moved transversely outward by perhaps 10
light years!   And often a new dot was seen near the core.

Such behavior is reproducible without actual superluminal space motions in a
special geometry:  The dots have to be traveling along a line pointing nearly,
but not exactly, at the Earth, and doing so at $\sim0.99$ c! The effect has
nothing to do with relativity; it's simply that the light travel time of the
dots for the second epoch was almost a year less, so the activity seems to be
speeded way up.

That hypothesis seemed extravagant to some, particularly because none of these
``blazars" could be interpreted as highly misaligned objects, which would be
seen to move much more slowly. Blandford and Rees (1978) \cite{Blandford:1978}
explained\footnote{*To be a bit more historically accurate, in 1978 there was
little direct imaging evidence for fast superluminal motion, and these authors
argued powerfully for the scenario using physical arguments based on data at
other wavelengths as well.} that the misaligned sources would, however, appear
much fainter (that {\it is} partly special relativity), and in fact were
identical to the tiny pc-scale relatively faint cores of the giant doubles.
After all, something must be feeding plasma and energy into those powerful
lobes. (In the early days, the jets (feeding tubes) were not clearly seen in
the images.)

You can see that in that case the superluminals would also have giant lobes,
one perhaps projected on the other.  When radio astronomers rudely pointed out
(e.g. after Blandford spoke at the famous Pittsburgh meeting in 1978) that that
was manifestly not the case, the theorists replied:  Get some dynamic range!
According to special relativity, those jets pointed at us are greatly boosted
in flux, neatly accounting for their very high incidence at high flux levels.

Because interferometer maps of the day were severely limited in dynamic range,
it was quite possible that rather bright lobes were present on arcsecond scales in association with
the pc-scale blazar cores. They could have been lost in the poorly subtracted sidelobes of the greatly
boosted compact jets. The boost can be very high because special relativistic
aberration directs the radiation into a solid angle of order the inverse square
of the bulk gamma factor of 10 or so; {\it plus} that factor (actually the
similarly-valued Doppler factor) comes in two more times, via photon arrival
rate and photon kinematic blueshift.

The low dynamic range was mostly the result of the ever-changing air column
densities over, for example, the 27 VLA dishes. This introduced 54 scalar
functions of time for the effects on amplitudes and (especially) on the phase delays
over each dish. Radio astronomers are by far the champions at mustering the
fortitude required to address data systematics with sophisticated
software.\footnote{*To this day X-ray astronomers mostly show
uncalibrated count spectra because they are too lazy to deconvolve their line
spread functions. They believe in X-ray
exceptionalism, so {\it only they are prevented from proper flux calibration}.
The published count spectra mainly show the
``effective area" curve, and diabolically conceal the physical information.
X-ray astronomers publish pages of these uninformative count spectra.}

It sounds bad that there are $\sim54$ unknown scalar functions required to correct
for the effect of the ever-changing atmospheric columns above each VLA dish.
However, the total number of measured fringe visibilities (baselines) is of order
$27^2$, not $2\times27$, per integration record. This brings massive redundancy
which can be exploited in favorable cases. The self-calibration algorithm was
being developed very rapidly around 1980, and it largely eliminated the
dynamic range limitation! Several groups began to look for diffuse emission
associated with the superluminals, including a major effort by myself and
J.~Ulvestad. We observed or reprocessed data on all the blazars\footnote{*The fast
Superluminal radio sources are approximately the same objects as optical
astronomers call blazars, based on their beamed synchrotron core emission which
can swamp even the optical spectrum of the AGN. I'm not making an explicit
distinction in this conversation-style condensed essay. That's not adequate for
research of course, and the subtleties are all discussed in e.g., Antonucci and Ulvestad
1985 and Wills \etal\ 1991, and my 1993 and 2012 reviews. The net effect
here however is to diminish work by others, e.g., Browne \etal\ 1983, which
used selection by VLBI speed. In general the present paper uses very incomplete and
self-serving referencing, for which the author expects to pay in Pergatory.
His major reviews of 1993 and 2012 are much better in this regard.}
known at this time, generally detecting the expected diffuse emission. With
strong arguments we convinced a lot of people of this Beaming Unified Model.
An important part of our paper was citing and debunking the various
counter-arguments in the air at the time. (That's the essential step shamefully
and consistently skipped by generations of accretion disk peddlers, as you'll
see below!)

\vspace{2.5mm}\hrule\vspace{2.5mm}\par

As an aside, it was fortunate that this self-calibration algorithm is a highly
iterative process. That was fortunate because it greatly prolonged the pleasure
of discovering the diffuse emission in each source. You have all sat in the
optometrist's chair while s/he interrogated you with questions such as: ``Now
which is clearer\dots image A\dots? \dots or B\dots?"  while slipping in or out,
say, a quarter of a diopter of correction.

The first several iterations of the self calibration map of very highly point
source dominated objects shows you nothing but that point source. If you do
reduce such data, be careful, because this program wants to please you. You
start with a point source model and the program tweaks the phases to give it
back to you. But if you aren't careful it will simply erase all the diffuse
emission, thinking that is what you want. Truly interferometry is an art,
and the optimal operation of these powerful complex nonlinear algorithms differs
depending on everything from source structure, to Fourier components sampled,
to phase stability over the antennas as a function of coherence length, etc.
It's a wonder it works at all. The only near-certainty is that you need some
SNR on every baseline time record or you will get garbage. Yet vast experience
speaks to the consistency and fidelity of the results in most cases, as long as
a skillful astronomer is driving the car.

\vspace{5mm}\hrule\vspace{5mm}\par

I'd started to boast above about my 1985 paper with Ulvestad.  I had occasion
recently to re-read that paper carefully, and had quite an odd feeling of pride,
but also the certain knowledge that 40 more years of practice have not made me
a better astronomer.

\vspace{5mm}\hrule\vspace{5mm}\par

Aside from general consistency of the detected diffuse emission matching
expectations for projected double lobes, we claimed we could {\it prove} the
idea was true! Our proof convinced at least the present writer. You might say
I underwent a religious conversion, and never since doubted its qualitative
correctness.

We said our {\it proof} has two axioms. First, the fluxes from the superluminal
cores are in fact greatly boosted in our direction. That largely follows from
special relativity and the linear superluminal VLBI sources, so no one wanted to
give up that axiom. The second was that the diffuse emission detected was
{\it not} highly directional. Most people were fine with that axiom as well.
It would be hard to imagine such giant diffuse clouds moving near light speed
and often the diffuse emission is two-sided. Optically thin synchrotron emission
from a multi-zone source is not intrinsically highly anisotropic. 
It hardly seems that {\it both} diffuse components would be
moving so relativistically in the {\it same} direction.

In most cases the diffuse emission we detected wasn't extremely bright because,
after all, the blazars were picked up mostly by their core flux. But several
had sufficient diffuse flux that those objects would have made it into our
low-frequency flux-limited catalogs {\it even without their bright cores}.

It would make no sense for all such objects to be pointing very nearly towards
Earth.  Of order a hundred times more would not be aimed so
perfectly.\footnote{At this time a theory paper by Blandford and Konigl
(1979) \protect{\cite{Blandford:1979}} made
me worry that these statistical arguments might be too na\"\i ve, due to a possible large
intrinsic beaming angle resulting from non-co-linear motions. But all the data
in hand suggested to me that if I did otherwise, I'd be too sophisticated.} So
for each object among the several referred to in the previous paragraph, there
should be $\sim100$ intrinsically similar objects pointed away.

According to our two hypotheses, misdirected equivalents of these blazars with
bright diffuse emission would have much weaker radio cores by Axiom 1; but the
second Axiom assures us that they'd have the same diffuse flux and would have
been included in the catalog anyway.

Since all these misdirected blazar equivalents are in the catalog, let's look
at what they could be.  The entire 3C list for example only has around 500
sources, and many are Galactic objects. The other entries are\dots simplifying
slightly\dots none other than the giant double quasars and radio galaxies! In
fact, we said ``many or most" of  the others would have to be misdirected blazars.    


All was not quite well in radio-astronomer-land however, as we shall see!

\subsection{b).  The Great Depolarization Asymmetry!  Another Proof of
beaming, but be careful what you wish for; Trouble with the cosmological
principle; Peter Barthel Feels Uncomfortable;}

Or, how the spectral Unified Model had to save the beaming Unified model,
and the two lived happily ever after. This is a story I told in
Antonucci 2012 \cite{Antonucci:2012}.  It would be pointless to rewrite it.

\vspace{5mm}\hrule\vspace{5mm}\par

Peter Barthel worked mostly on VLBI observations of superluminal sources in the
1980s, and he knew well that the beam model explained many properties such as
superluminal motions and jet sidedness qualitatively.  But he was (according
to the title of his rumination for a conference \cite{Barthel:1987}) ``Feeling Uncomfortable"
because one had to assume that a paradoxically large fraction of these sources
in quasar samples have jet axes fortuitously close to the line of
sight.\footnote{Note though that he did not entitle his paper ``Do all radio
galaxies contain hidden quasars?" He did not make that superficially similar
statement because he is a very smart guy. The answer to that question would be
no, but the answer to the question he posed is still basically yes.}
Barthel later wrote a famous paper (Barthel, 1989 \cite{Barthel:1989}) entitled ``Are all
quasars beamed?" suggesting that those quasars whose axes lie near the sky plane
somehow fall out of quasar samples, and (inspired by the spectropolarimetry)
in fact get classified as radio galaxies.

The general idea of beaming to explain superluminal motions and one-sided jets
was accepted by most doubters as a result of two key back-to-back discovery
papers in Nature, reporting on the so-called Lobe Depolarization Asymmetry.
They are Laing (1988) \cite{Laing:1988} and Garrington \etal\ (1988) \cite{Garrington:1988}.
It was spectacular, inspired work.
The wonderful astronomer and human being Peter Scheuer (1987) \cite{Scheuer:1987} described it in a
section of a (basically unavailable) conference paper called, ``Why
relativistic beaming is true." (We'd know a lot more about radio sources today
if Peter had been granted a reasonable time on Earth.)

Giant double radio quasars are rather similar to giant double radio galaxies,
but there are differences. They tend to show one-sided, fairly bright
arc-second-scale ($>$kpc) jets. Radio galaxy jets are hard to see and, when you
can detect them, they are more similar in flux on the two sides.
Both the Laing (1988) paper and the Garrington \etal\ (1988) paper studied samples of
double-lobed {\it quasars}.

Some extremely clever person designed the VLA to work in scaled configurations.
That is, the standard radio bands are separated by factors of three in frequency.
The 27 antennas are laid out along railroad tracks, and periodically moved
outward en-masse by factors of three in distance from the center. So to make
images of spectral slopes or depolarization one can use for example the 20~cm
band in the largest configuration ($\sim30$~km), and compare the result to mapping
at a frequency in the 6~cm band with a 1/3 scale array! Of course with
interferometry we just measure a finite sample of Fourier components of the
image brightness distribution and only those components inform what you see in
the image plane. With this plan the maps at the two frequencies would be based
on the same Fourier components.
(The {\it clean} algorithm tries to guess at what the unmeasured
components would have been.)

The following sounds implausible not only because of the powerful result but
because of various coincidences too weird to explain.  These papers showed that
in the 6~cm band observed with the more compact array configuration, in
{\it all} cases I think, both lobes were shown to be highly polarized.

Sit down for this, though. In all, or perhaps nearly all cases, the 20~cm images
showed that only {\it one} lobe is highly polarized in each quasar. It was
always the ``jetted" side, and the nearly inevitable interpretation was that
this proved the jetted side is in fact the near side (a commonplace now\dots).
It lies in the foreground of some polarization-angle scrambling medium, while
the far side was almost completely {\it depolarized}, undoubtedly due to Faraday
rotation measure gradients within each synthesized beam (angular resolution
element). Aside from the breakthrough information this suggests to me that the
diety or dieties possess a very jolly sense of humor to put the effect exactly
in the parameter space observed; and in virtually every case that s/he/it/they designed the
sources with the VLA in mind, to delight only us on Terra. I can't prove it but
I have heard that the former Astronomer Royal warned us that it would be really
funny if nothing funny ever happened. But there is a limit to my credulity.

To recap, and blend in something else that delighted me:
There is a very strong tendency for one radio lobe in double-lobed radio
{\it quasars} to be depolarized at low frequencies (say the VLA A Configuration at
20~cm) by Faraday rotation within the observing beam\dots but {\it only} on the
sides of the (single-jet) sources which {\it lack} the jet (Laing 1988;
Garrington \etal\ 1988) \cite{Laing:1988,Garrington:1988}!
Most people accepted that the depolarized lobe must be the more distant one,
located behind a large-scale depolarizing magneto-ionic medium; thus the
polarized lobe is on the near side, so that the jet is also on the near side,
{\it just as expected} for the beam model!!

Church bells rang. But be careful what you wish for!  The effect was way
too strong! The first paper of the pair has this disclaimer: ``The sources
observed here must then be oriented within about $45^\circ$ of the line of
sight\dots to generate sufficient asymmetry in path length\dots" to fit the
depolarization data!


{\it Neither these authors nor the referees} must have been particularly curious
people, not to demand elaboration!\footnote{*The selection criteria in Laing (1988) \cite{Laing:1988} and in
Garrington \etal\ (1988) \cite{Garrington:1988} slightly
but insufficiently favored low-inclination sources. Ulvestad and I didn't have
this problem because of course we did know about the polarimetry, and we didn't
``know" that quasars should be considered separately from radio galaxies. Too
much ``knowledge" is as dangerous as too little.}
They did not ask themselves even
rhetorically what the hell happened to the quasars oriented close to the sky
plane!
They were apparently blissfully unaware that of course we knew from optical
spectropolarimetry that many of the high-inclination ``quasars" are simply
masquerading as FR II radio galaxies, which would have dropped out of samples
restricted to quasars, as explained by Barthel 1989 \cite{Barthel:1989}.
Barthel had read outside his field, and found our polarization papers. There
are many mysteries in one's field which one simply can't resolve without
reading outside it.


\section{Part 3. The peculiar AGN Reddening law}

Using only hard data, special relativity and the Copernican principle, we
derived an extinction curve for radio quasars. We claim it's the most robust
determination of the law and mention some consequences, including the equally
robust derived anisotropy as a function of wavelength throughout the electromagnetic spectrum.
We need also to reassess the broad Hydrogen line spectra, in particular the very large
Balmer to Ly $\alpha$ ratios. This type of extinction totally dominates that of
most radio loud objects, and plays a major role in most quiet ones which, however,
often also have more standard reddening from dust outside the nucleus as well.

\begin{figure}[htbp]
\includegraphics[height=8.5cm]{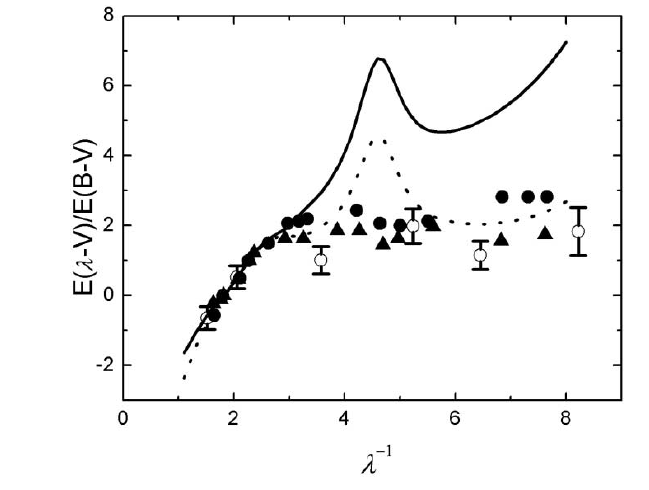}
\caption{Reddening curves from Gaskell \etal\ (2004) for composite
spectra of various values of radio-core dominance. Filled
triangles are from comparing ${\cal R}\ge1$ with ${\cal R}\le0.1$,
and filled circles are from comparing $0.1\le{\cal R}<1$
with CSS. The open circles represent average BLR extinction
values between face-on $({\cal R}\ge1)$ and edge-on
(${\cal R}<0.1$; CSS) objects. Theoretical reddening curves
derived from CCM89 for $R_V=5.3$ and 3.1 are shown as dashed
and solid curves, respectively.\label{fig:6}}
\end{figure}

One can make a robust extinction curve by dividing the SEDs 
of objects of different extinction, {\it if the comparison is between sources known
to be intrinsically identical}. That's the key feature of the curve put forth here
in Fig.~\ref{fig:6} from Gaskell \etal\ (2004) \cite{Gaskell:2004}. You needn't
believe my conclusion but I will specify what axioms you have to give up in
that case.

\vspace{5mm}\hrule\vspace{5mm}\par

We reported (Gaskell \etal\ 2004) on a study which was so well-designed, its
dramatic result depended logically only on hard data, special relativity, and
the Copernican principle. We studied a {\it complete lobe-selected} quasar sample.
The fluxes from the radio lobes are manifestly at least approximately isotropic.
In that situation (aside from small statistical effects), if a giant reorients
a member of the sample it will have the same lobe flux and necessarily look like
another member of the sample. {\it The power of that idea is shown also by the very strong
conclusions reached for High z 3Cs by Marin and Antonucci (2016)} \cite{Marin:2016}.

In the 2004 sample there was of course a wide range in core dominance. Quasars
with strong radio cores show, to first order, linear superluminal jets. Special
Relativity says the fluxes from such cores are highly boosted in our direction.
The Copernican principle says there are many other quasars not so specially
oriented. They {\it must} comprise the sample members with lower core dominance.

Thus to determine the polar diagram or anisotropy or extinction curve of any
component, one need only divide the ones with weak cores by the ones with
strong cores.
The result is an extinction curve that is quite selective in the optical region, and quite flat in the
near-UV. Small grains are evidently severely depleted. 


Then we tested that curve as follows. Due to the wonderful work of Joann Baker,
with advisor Richard Hunstead in a supporting role, we had up to 6 measured
emission lines, going from Ly $\alpha$ to H $\alpha$ for every quasar. We divided those
in the same way, according to core dominance to derive an independent {\it line} extinction curve.

{\it It was identical in shape and normalization to the continuum curve.}

Most reddening curves are only suggestive because there is no assurance the
comparison objects are intrinsically identical.
Since our papers, authors that cite reddening curves in the literature list
ours among the others, all of which are of the latter type. There is no
recognition that ours is the robust definitive one. In fact some people
perceive that e.g. the SDSS curve supercedes or even disproves ours.
Such is the level of scholarship in this field.


Since then our reddening curve has proven its plausibility and its worth.

I've come to realize that the AGN torus is almost certainly a
{\it winnowing machine}. Every grain has a different sublimation radius and
physics tells us only the large grains, being better radiators, reach and
define the inner edge of the dusty torus. So I don't consider the dearth of
small grains required for our extinction curve to be a stretch; just the opposite.

There is another qualitative point of consistency. A major result from
mid-infrared interferometry is that one tends not to resolve the tori in the
equatorial direction, which is as expected for a pc-scale dusty gas with Av
$\sim1000$ (Krolik and Begelman 1986, 1988) \cite{Krolik:1986,Krolik:1988}.
More often one resolves the Narrow
Line Region in the polar direction (e.g. Bock \etal\ (2000) \cite{Bock:2000}, which didn't require
interferometry for NGC1068). Although this emission is optically thin, it lacks
the usual Silicate emission features due to small Silicate grains. That suggests
the dust has passed through the torus crucible, losing small grains and perhaps
the Si grains generally \cite{Ramos:2017}.


It was known to the ancients that the ratios of Ba line strengths to
Ly $\alpha$ are far above case B, and that only a curve shaped like ours could
account for it and still have mostly the physics of Case B.
Martin Gaskell \etal\ (2017) \cite{Gaskell:2017} have
shown how well this works for the integrated line fluxes. They argue that the
wings, which are said to be energetically insignificant, can be explained by a
small Case C contribution. This is my optimistic broad-brush description, but
there are many details about line profiles and transfer functions which must be
considered. What I can defend is this: our extinction curve is manifest in the
continuum and line spectra of AGN, and changes both significantly. Correction
greatly reduces the anomalously high Balmer to Ly $\alpha$ ratios. A piece of
the puzzle which has yet to be rigorously incorporated is the energy budget and
the infrared dust emission.

\vspace{5mm}\hrule\vspace{5mm}\par

\section{Part 4. The Tragi-Comic Story of AGN Accretion Disk Modeling.}

It's been absolutely routine for the past 40 years to assert that accretion disk
models can account for the optical/UV continua of AGN. In a recent example in
Science magazine, C J Burke \etal\ (2021) \cite{Burke:2021} open their article as
follows: ``Accretion disks are present around growing supermassive black holes
(SMBHs) found in active galactic nuclei (AGNs). Standard theory of
radiatively-efficient accretion disks can reproduce the broad-band emission
from AGNs ..."\footnote{We will see below that the disk model or anything
resembling it is completely at odds with the spectral energy distribution.}


Yet one of those authors distinguished by intellectual honesty, when writing
alone, states in a public funding application, 
``In marked contrast to models of accretion disks around stellar mass black
holes, neutron stars, and in cataclysmic variables, existing theoretical models
of accretion disks around supermassive black holes do a very poor job of
explaining, never mind predicting, the observed properties of luminous
active galactic nuclei (AGNs) \cite{Blaes:2021}."


What's going on here? The prestigious Science paper with the same author says
the opposite. It also very strongly implies that with the model tweak presented,
the theory applies to AGN. There is no mention of the counterarguments such as
the factor of 10 too low surface brightness measured by microlensing (e.g.
Dexter and Agol 2011) \cite{Dexter:2011}.


The Science paper purports to address just one of the many contradictions, and
fails at that, though they do not say so. It attempts to explain the extremely
rapid variability, but via an opacity/convection argument that applies only
locally in the disk, which doesn't help at all since the ``problem" appears
equally throughout the entire continuum supposedly produced by the model,
covering several orders of magnitude in frequency. Finally, see their Figure~1,
proposed as compelling, but in fact a terrible fit. Such is the level of
scholarship in my field. This fancy paper is disingenuous in my opinion.

Perhaps the most fundamental prediction of the standard disk model is that the
characteristic temperature is proportional to the one fourth power of $L$, and
the inverse square root of $M$.  We have methods of estimating $M$ and we can also
get the approximate BBB luminosity from observation.  These scaling relations
have been tested implicitly in many papers and explicitly in at least three:
Courvoisier and Clavel 1991 \cite{Courvoisier:1991},
Bonning \etal\ 2007 \cite{Bonning:2007} and Davis \etal\ 2007 \cite{Davis:2007}.
See also Reimers (1989) \cite{Reimers:1989} No sign of the predicted behavior is seen! 

These papers had no noticeable effect on the consensus however!

I note that when the Courvoisier and collaborators debunked the disk model,
they followed up with several very nice modeling papers based on
generic shock cascades, which is consistent with the variability as well.  




\vspace{5mm}\hrule\vspace{5mm}

Let's return to the BBB spectral energy distribution.  

\begin{enumerate}
  \item Of course we can't generally observe much past 1 Rydberg, and when we can,
the data are affected by intergalactic absorption.  
  \item Wavelength-dependent extinction is common as shown above for radio loud
quasars, and this can't be corrected reliably in individual objects.
  \item Often the host galaxy contributes and it is difficult to subtract it
unambiguously.  
\end{enumerate}

All of these problems are avoided by the T-shirt method described below!

Let's back up to my start in science. 
When I started out in the 1980s, I realized that quasar continuum spectra carry
very little interpretable information. They are just rather generic looking
broad convex lumps which one can fit by many functional forms with those properties.
Many authors had already noted their generic nature, perhaps starting with
Ward \etal\ (1987) \cite{Ward:1987}.



Following the insights of Lynden-Bell (1969) \cite{Lynden-Bell:1969}, most workers agreed that to
provide the luminosities observed in quasars, and the energies contained in
radio lobes, black hole accretion was the only game in town. It was well-said,
probably by M.~Rees, that accreting supermassive black holes are by far the least
amazing idea that fit the facts.

Following the suggestion of Shields (1978) \cite{Shields:1978}, this idea was instantiated in the
simplest possible way as quasistatic geometrically thin, optically thick
accretion disks. The potential gradient was used to predict the temperature as
a function of radius, and thus the SED. Soon as pioneered by
M.~Malkan (1983) \cite{Malkan:1983},
models incorporating these ideas were fit to observations of AGN.

Since the get-go, this paradigm required major cognitive dissonance. 
The quasistatic assumption is crucial to the models, but that was known to be
comically violated by the speed and extremely broadband nature of their flux
variations. The discrepancy is by several orders of magnitude, as discussed
for example, by Alloin \etal\ (1985) \cite{Alloin:1985}. Those authors explained that their
observations of NGC1566 showed that the H$\beta$ emission line and the optical
continuum vary on timescales of days, with a very tight upper limit on
any temporal offset between the two time series. That means the ionizing
continuum and the optical continuum vary extremely rapidly and closely in
phase, utterly inconsistent with any viscous quasistatic flow.


Also in the 1980s, I pointed out that these models necessarily produce
{\it significant polarization in the disk plane}. Since the radio jets
(including the little ones in Seyferts) lie parallel to the polarization,
essentially every thin disk model in the mighty river of such models that
continues to flow today has as an essential feature: {\it the requirement that
the jets emerge from the sides of the disks!!!} Such is the level of
scholarship in my chosen field.

I discuss these and other horrible and palpable contradictions in Antonucci
\etal\ (1989) \cite{Antonucci:1989}, and many papers since. Almost no one cares.


Those of you who had the pleasure of knowing the late, great Bill Mathews of
UC Santa Cruz will appreciate this story. I was complaining to him that this
obviously erroneous model was accepted instantly and nearly universally, and
told him about my commentaries and ongoing tests.  He obviously agreed, but
cut me off soon with the interjection:. ``Oh you can't stop that!"

The main topic of that 1989 paper was a search for the Lyman continuum
absorption feature predicted for the standard disk by Kolykhalov and
Sunyaev (1984) \cite{Kolykhalov:1984}.

In a carefully designed plan, we showed that quasars {\it lack that
feature}\footnote{*As we explain, the lack of {\it any} feature in quasar spectra at the Lyman edge
position has other remarkable implications. Consider that the Broad emission
Line Region could very reasonably be expected to absorb ionizing photons, and
to produce Lyman continuum emission. Neither such effect has been documented
even once to my knowledge, while they would seem to occur fairly generically.}
\footnote{** After our paper was published, another paper claimed to see absorption in
the Lyman continuum consistent with an origin in a disk atmosphere in a couple
quasars. Those authors carelessly omitted a key test to distinguish edges from
disk atmospheres from ``associated" Lyman Limit Systems from foreground gas:
kinematic and gravitational broadening would completely disguise the
accompanying narrow absorption lines that a cloud would produce. We used that
test to rule out the disk explanation for our candidates. Then Kinney and I
later checked these others and alas they all showed the accompanying narrow
absorption lines.}

Blackbody accretion disks also predict a spectral slope of 1/3 over a wide
frequency range with slightly more realistic physics. The spectral index remains
positive.
Since almost all quasars actually have
{\it negative} spectral indices, Malkan fit the models not to the observed
spectra but to the residual spectra {\it after subtracting a notional steep power
law extrapolated from the near infrared}. But universal minima at $1\mu$ and
reverberation mapping show that the near-IR is dust reradiation of the primary
continuum, which must drop like a stone shortward of $1\mu$. The community was
convinced of this largely by a famous paper by Barvainis (1987) \cite{Barvainis:1987};
see also Clavel \etal\ 1989 \cite{Clavel:1989}. Thus the extrapolation under the
optical was\dots not legitimate.  



Malkan also got about the same temperature for each object, evidently because
he was interpreting an atomic feature as the exponential cutoff expected at the
innermost stable circular orbit.

The next attempt I studied closely was by A Laor, and I found it very
instructive in multiple ways.  He fit a larger sample, crucially finding it
advantageous to assume maximal spin. He concluded that the majority fit the
models in the $\chi^2$ sense. It was an odd criterion because I'm sure,
having examined a solar spectrum with great intrinsic SNR and high resolution,
that I could use the method to rule out the G star model.


A cursory look at
the analysis confirmed that the quasars that fit the criteria were basically
the ones with low SNR. Still, I wondered how the luminous objects, necessarily
with high mass and thus low temperature, managed to pass muster.  As I read, I
predicted silently and quickly confirmed that they were strongly biased
at very high inclination to the line of sight. That combination of maximal spin
and near $90^\circ$ inclination results in beaming highly blueshifted light
towards the observer! Thus the quasistatic disk model was saved at the wee
cost of the Copernican principle. It was slightly worse than that in the sense
that if aligned with the matter accreting through the AGN torus, they would alas
have been observed as Type 2 objects and not quasars.

Thus I saw that the lack of any sign of the expected relation
$T\propto L^{1/4} M^{-1/2}$ could easily be subsumed into the many nuisance
parameters like spin, inclination, host subtraction, reddening etc, not to
mention the ever-ready designer Comptonization. 

It now gets a lot worse.  Laor and colleagues Netzer and Piran were apparently
aware that the polarization of quasars was inconsistent with the high disk values
based on the expected scattering dominated atmospheres. But only that the
observed polarization was too low, and not that it was in the wrong direction!

They adjusted the atmospheric densities such that the optically emitting annuli
would be absorption dominated, and produce low polarization, which would then
rise strongly into the UV, which would be dominated by hotter annuli.

Malkan had noticed the sadly low observed \% polarization too, and argued that the unified
model censorship of edge-on quasars helped a lot by removing those which would
have had the highest \% polarization; then the magnitude distribution would supposedly be
as predicted by the disk model.

None of these authors seemed to be aware that as I'd pointed out in papers in
the 1980s, the polarization was in the wrong sense for a disk atmosphere, and
they were thus wedded to the prediction that the jets emerge from the sides of
the disks. That remains true of models being published today, though theorists
don't mention it. Some are aware of it because I bring it to their attention,
but do not deem it worthy of mention in their papers. A random example is
Mitchell \etal\ 2022 \cite{Mitchell:2022}. These particular authors\footnote{I'm
glad these authors have seen the light but astounded that they don't acknowledge
the previous 35 years of robust demurs. The reader would truly think it's a new
discovery. Not a hint is given that a huge superset of this model was disproven
many times over many decades.} comically crow that they've
discovered some wee problem with their quasistatic disk theory, but hide the
fact that much more powerful arguments have been published and spoken about at
meetings for 40 years, which falsify a huge superset of models, including
theirs. That style isn't honest.

I knew Joe Miller understood the situation with regard to the erroneous sign
of polarization of disk models, so I said to him (with tongue in cheek)
``We know the polarization has the wrong sign, but all these authors have shown
that it has the correct magnitude."

I knew his sardonic facial expression by that time. He just shook his head and
said, ``That doesn't help when you're balancing your check book."

Digression: It pays to cheat! Truth is a legitimate defense for libel. I ask
my readers if they can top the following story for brazenness and mendacity.
First I noted in print (Antonucci 1988; Antonucci \etal\ 1989 \cite{Antonucci:1988,Antonucci:1989}) that the
polarization direction in most quasars and other Type 1 objects, being parallel
rather than perpendicular to the axis measured in the radio, negates the disk
model. Next, Laor \etal\ (1990) \cite{Laor:1990}, used the parallel polarizations
as though they were perpendicular, an egregious error negating their whole study.


Next, I pointed out their error in 8 papers starting with Antonucci 1993 \cite{Antonucci:1993}
and including Kishimoto \etal\ (2004) \cite{Kishimoto:2004}.

Can you imagine my consternation and astonishment to read this recently in
Capetti, Laor, Baldi, Robinson, and Marconi (2021) \cite{Capetti:2021}
``In contrast with our conclusion that the continuum and
BLR polarization are produced by a single scattering medium,
Kishimoto \etal. (2004) suggest that the optical polarization in
quasars is produced by electron scattering within the accretion
disk atmosphere\dots However, this interpretation is inconsistent with the PA
of a disk atmosphere scattering, which is predicted to be perpendicular to the
radio axis, rather than parallel, as observed."

They literally reversed history, stated that we'd made the error, and
pretended that they were correcting us! 

Wait, there's more. That same paper misrepresents the work I described in two
{\it equally epic} ways.   

They added two additional inane arguments, attributed them to our paper, and
refuted them as well! One stated a completely fallacious reason for our
attribution of the polarization to electron scattering, and another did the
same for our identification of the Ba edge feature. The arguments they put in
our mouths were not only inane, but made up out of whole cloth.  
An astounding trifecta in the field of calumny!!!

When we expressed our dismay and asked for a correction they simply said no.
So they win. I suppose any interested reader wouldn't bother with our ``true
central engine" papers, after reading all that, and would perhaps share a
chuckle with those worthies, take my group down a peg in their estimation,
and miss our best ever result. Or so goes my feverish mind.

I don't know what to make of it. Far fewer people will read this obscure
rebuttal so I suppose I'll be thinking of those guys as my spirit leaves my body.

\section{Part 5. The Naked Central Engine Spectrum of Quasars.}

The only 5 spectra you've ever seen of the quasar central-engine.

Polarimetry has allowed us to objectively remove all atomic and dust emission
to reveal the true central engine spectra from $0.3\mu$ to $1.6\mu$ in the rest
frame. It attracted virtually non-interest.

This is probably the best thing I've done, and I say that with the disclaimer
that actually my then-postdoc Makoto Kishimoto did virtually everything. Thus
this is not a boast.


Considering the fiasco of first principles theory, I thought a more
phenomenological approach was in order both for observations and theory.

I thought a good starting point was to check the assumption of optically thick
thermal emission. We know stars are optically thick, with higher temperatures
inside, because wavelength of high opacity are seen in absorption. For quasars
in which mildly relativistic motions and gravitational redshifts are expected,
line features would be smeared out. Therefore we went after the Lyman continuum
absorption feature, which was indeed predicted to be deeply in absorption
(Kolykhalov and Sunyaev 1984) \cite{Kolykhalov:1984}. Theirs wasn't a definitive
calculation but a combination of results from the available stellar atmospheres
with the lowest surface gravity.


Alas, quasar show no features of any kind at all at that location! We designed
a powerful series of observations to maximize sensitivity to highly broadened
edges using 8" slit widths for accurate photometry. We also got high resolution
spectra of the edge positions to be sure to distinguish smoothed atmospheric
edges from associated Lyman Limit Systems, and also of the Lyman~$\alpha$ region
because a true atmospheric edge would not produce a strong sharp Lyman
absorption line. Later attempts by others failed to take these precautions
and, unbeknownst to those authors, produced only false positives.

It's possible that Lyman edges, arising deep in the potential well, are
inconspicuous because of heavy scattering and the disks are famously unstable
in the inner annuli where they arise.
While we designed our observations to detect even quite broad features that
could decrease the contrast excessively, at least in principle.
Thus we decided to check the behavior at the Ba edge position, which arises
from much more distant annuli, alleviating all these concerns.

My philosophy throughout decades of testing the standard and non-standard disk
models has been to give the model the benefit of every doubt, bacause it seems
so attractive and natural.
That would seem impossible to check for Ba continuum absorption to most people
because at that very wavelength, one observes enormous Ba continuum emission
from the broad line region. But perhaps the magic of polarization could save
the day.

\vspace{5mm}\hrule\vspace{5mm}\par

Digression into the lack of Lyman edge features in quasars.

It'd be worthwhile even today to read the discussion from our 1989 paper on the Lyman
Limit region, because other mysteries are discussed, too. For example, every quasar
shows enormous Ba continuum from the BLR.  An epic number of free electrons
combine to the $n = 2$ orbital.

However, {\it no electrons in any object} detectably combine directly to $n = 1$!!!
That jump has a much higher capture cross-section. What do you think of that?
Optical depth effects can weaken the Lyman emission feature, but
extensive experimentation fails to reveal any modeled cases where the continua are as
perfectly smooth across 912~\AA\ as any quasars.\footnote{*This information
comes to me mostly from a lifetime of challenging Gary Ferland to make Cloudy
models that look like quasars in this part of the spectrum.}

It gets worse. Photon counting arguments lead to a covering factor for
recombination-line producing BLR clouds of at least 30\%, if the Ly~$\alpha$ line
or the Balmer lines have strong contributions from that process. (The latter
are much stronger relative to Ly~$\alpha$ than for Case B recombinations, so some
other processes probably occur, and non-standard reddening also plays a major
role: Gaskell \etal\ 2004; Gaskell \etal\ 2017 \cite{Gaskell:2004,Gaskell:2017}.)
Nevertheless a substantial covering factor in
clouds opaque at the Lyman edge is required. If such clouds were larger than the
1 Ryd-emitting continuum region at least a large minority of quasars should show
spectral cutoffs there. If instead the clouds comprise a fine mist composite
spectra should show a robust partial absorption at 912~\AA. Yet of the
countless thousands of edge locations examined, I am unaware of a single
intrinsic Lyman continuum absorption edge which can be identified with a BLR
cloud. What do you think of that?

I can only assume that those objects in which our sight line passes through
recombination-line producing clouds\dots drop out of the sample and aren't
called Type 1 AGN! A moment's reflection identifies the only candidate for
their observed properties: they must be classified as Type 2. That is, every such sight line must also
pass through the torus. Thus as seen from the nucleus, the solid angle covered
by recombination clouds must be a subset of that covered by the torus. Various
colleagues have convinced me that the outer boundary of the recombination BLR
is continuous with the inner boundary of the torus, largely from the
reverberation transfer functions. But ionizing photons incident on the torus
would produce recombination lines. I can only conclude that these BLR clouds are
simply {\it the name we give to the part of the torus which extends inside the
sublimation radius.}
This is all discussed in Antonucci \etal\ 1989 \cite{Antonucci:1989}, with
some further considerations in Maiolino \etal\ 2001 \cite{Maiolino:2001}.




We now return to the test of the behavior of the quasar {\it central engine}
continuum at the Balmer edge position.
Quasars generally tend to have a slight polarization which is parallel to the
radio symmetry axis which is wavelength-independent (Antonucci 1988;
Schmidt and Smith 2000; Kishimoto \etal\ 2003, 2004, 2008 \cite{Antonucci:1988,
Schmidt:2000,Kishimoto:2003,Kishimoto:2004,Kishimoto:2008}).
The broad line polarizations can be similar, or somewhat different with
structure inside the profiles. We sought and found 5 examples in which no
polarization at all was detected in the lines. The slight $\ltwid1$\% polarization
in the continuum is exactly wavelength-independent to the attainable accuracy.
We propose and then show
that this slight polarization is due to scattering. Then there are two powerful
arguments that the scattering is by free electrons rather than dust grains. The
first is the precise wavelength-independence of \% polarization in the continuum, so that the continuum
polarized flux is identical in shape to the total flux. The second is that the
scattering takes place interior to the broad line region, since the broad lines are
unpolarized. That places it {\it well} inside the sublimation radius, so by
definition, {\it in a dust-free region}.

Again we sought and found these five precious quasars, and by design their
redshift put the desired Ba limit wavelength at the most convenient location
in the observed frame; and which happened to have the slight polarization
{\it in the continuum only}.

The polarized flux spectra are nothing but a noisier version of the total flux,
{\it but with the emission lines and bound-free continua magically removed!!}
See Kishimoto \etal\ (2003, 2004) \cite{Kishimoto:2003,Kishimoto:2004}.

Thus we produced the first ever spectra of the quasar central engines,
isolated from all contaminating atomic emission!   Virtually no one showed
interest in these previous and unprecedented and unique spectra!
We found that in all five cases the central engine spectra show the Balmer
continuum in absorption. That is, they somewhat resemble the spectra of A stars!
This is an extremely important result and proves that at least some of the
central engine optical continuum arises in optically thick material with a
normal temperature (or source function) gradient. That is, hotter inside!

Later we performed the same trick to find out what the central engine spectrum
does in the near-IR, heretofore completely inaccessible underneath the dust
emission which is ubiquitous longward of $1\mu$ in the rest frame.
Again, all five quasars behave the same way: A slope change occurs so that,
whereas the spectral index of a fit to $F_\nu\propto\nu^{\alpha}$ is almost
always negative in the optical/UV, in the near IR it breaks to $+0.35\pm 0.10$. 

These are arguably the only known spectral features in the central engine spectrum and
constitute a major advance, but it attracted virtually no community interest.   
Together with all the other constraints (such as microlensing surface) it tells
us a lot about the emitter, that at least part of the light comes from optically
thick gas (with heat dissipation at large optical depth) and effectively only
partial covering (with dark regions in between the bright regions).

This is about all that I know about the quasar optical/UV continuum today. 
The nearly generic shape of the Big Blue Bump; the two localized features
just described, the Ba Continuum absorption and the Near-IR turnover; and
the grossly sub-blackbody surface brightness from microlensing. There is also
sometimes a slope change near 1000~\AA\ as well, though that is probably due to
absorption by wind.

\vspace{5mm}\hrule\vspace{5mm}\par

A way forward?

There is however a clue from a generic feature of the AGN SEDs. We recognized
the emission mechanism for the infrared light in part by the fact that it has a
feature (really two!) which is fixed in wavelength. Virtually every total-flux
spectrum has a conspicuous minimum when plotted in energy ($\nu F_\nu$) vs.\ log frequency
at $1\mu$. Thus the IR emission cuts off sharply of $1\mu$. This is just
what one would expect for dust emission, because the most abundant dust species
sublimate at around 1500~K. (Similarly, though not quite so precisely, the dust
bumps terminate in the sub-mm region.) That's also quite understandable because
dust anywhere in the host galaxy is heated to a few 10s of degrees.

What is the analogy for the Big Blue Bump? Until the polarization work
described above, the low-frequency turnover was an open question because the BBB
couldn't be traced past $1\mu$ due to the contaminating dust emission. However
as explained above, we now know that the optical slope, approximately
$\alpha = -0.3$ or so, breaks\footnote{This break can't be identified with the so-called self-gravity radius of an
accretion disk, which would be at longer wavelengths. In fact it's not clear how
such a cutoff would arise in the accretion process, though self-gravity breakup
would make sense if the disk ran backwards! Gas can break up due to
self-gravity, but gravitating objects like stars would not obviously decompose
and turn into a gas disk on the way in! See Goodman 2003 \protect\cite{Goodman:2003}.}
to +0.35 in the near-IR. Thus the BBB becomes
energetically insignificant past that region (the $\nu F_\nu$ slope being +1.35).

So it's fair to say that the BBB begins around the same location that the dust
emission ends. (The very narrow range of wavelength for the $\nu F_\nu$ minimum is also
suggestive of that, since otherwise the drop in the dust spectrum might not
define it quite so well.)

Where does the BBB end at high frequencies? The answer to that is very
remarkable as well because it is closely consistent from object to object.
One generally observes, from the X-ray point of view, a Soft X-ray Excess:
that is, the flux below $\sim1$~keV in energy is usually well above the
extrapolation from higher energy. If fit with a blackbody, whether or not
physically appropriate, this excess gives a temperature of  $x\sim10^6$~K, with
very little dispersion.

Thus the BBB extends primarily from
$1\mu$ ($\sim1$~eV*) to a break at say 300~eV. Such behavior strongly suggests
 the controlling element is atomic physics rather than global conditions. Why
this range is generic is not known at present. Of course efforts have been made
to use this valuable clue, none totally successful so far.

Rapid broadband variability is a key property of AGN and it's wise to make that
part of the model rather than making the false and manifestly disastrous
quasistatic assumption! To support the role of shocks, I note that Tidal
Disruption Events are precious examples of supermassive black hole accretion,
and shocks are very likely essential there. Note these two key references,
simulating the prompt emission of TDEs: Huang \etal\ (2023) \cite{Huang:sub2023} and
Ryu \etal\ (submitted, 2023) \cite{Ryu:sub2023}.
%
%
See Courvoisier and T\"urler (2005) \cite{Courvoisier:2005} for a more qualitative
description of how shock cascades might account for all the accretion radiation.

\section{Part 6. Disproof of the accretion disk and related models, suitable for
illustration on a T-shirt. How to beat a dead horse.}

I've written many papers over the decades pointing out robust falsifications of
the disk model, and in this enterprise I've been largely alone and largely
ignored.\footnote{An important exception is the work of Courvoisier and
collaborators, starting with Courvoisier and Clavel (1991) \protect\cite{Courvoisier:1991}.
These strong papers also attracted no attention.}
I'm sure this won't change with this T-shirt argument, which I've also
given before. Yet I claim it can disprove any model involving optically thick
thermal emission from a fixed area, a huge superset of the standard disk models.  
This will be presented here in qualitative form in the hope that a more
industrious person carries it out.
Nevertheless if you are well acquainted with the data, you too may find it
striking.

My hot dust lag argument is extremely valuable because it avoids all the
limitations of direct observation of the BBB, enumerated in Part 4!  
True, it doesn't tell you the shape of the BBB. But it tells you something
incredibly valuable, it covers the entire BBB and it is immune to reddening
and host galaxy contamination! It tells you that {\it from the point of view of the
innermost surviving dust grains} the shape is extremely generic and thus independent
of $M$ and $L$ and anything else!
The standard disk model, and any optically thick thermal model in which the
radiating area is proportional to $M^2$, would not have that property!

Many excellent papers have presented reverberation distances between the BBB
source and the hottest dust emission. The distances follow the expected
$R\propto L_{\rm BBB}^{1/2}$ just as one would expect from radiative equilibrium and
sublimation. {\it But the dispersion in the relation is interestingly low}.
The relation is so tight that this correlation provides an excellent standard
candle for cosmology!  See the VEILS project simulation in H\"onig \etal\ (2017) \cite{Hoenig:2017}.

In Koshida \etal\ (2014) \cite{Koshida:2014}, which covers luminosity range of around a factor of 300
in $L_\nu$ (their Fig.~10), they wrote a dispersion in hot dust lags in K band of 0.13
in the log. They assess the measurement uncertainty at 0.11, and note that there
are several other sources of dispersion which must be present. Thus {\it no
intrinsic dispersion} is detected, and they set a very generous upper limit at
0.10. This is indeed conservative because subtracting 0.11 in quadrature leaves
0.07. That's still not the dispersion in the bolometric correction from $M_V$ to
$L_{\rm BBB}$ because that remaining error budget must also accommodate those
introduced from V band variability, anisotropy of the BBB emission, extinction,
host correction, geometry and inclination of the torus, etc. In
the disk model, the anisotropy alone would easily contribute that much I think,
leaving none (or less than none) for the effects of T.

The next paper in the series, Minezaki \etal\ (2019) \cite{Minezaki:2019} extends the luminosity
range to four powers of ten (!) without a noticeable increase in dispersion, but no value
is quoted there. Their Fig.~5 is suitable for a T-shirt. But they also show
that using mid-IR or X-ray instead of V band (less susceptible to anisotropy,
reddening, host subtraction, etc.) tightens the noose!


That's very remarkable and has two implications. First, the innermost, hottest
dust has extremely similar physical properties from object to object. The
sublimation temperature is very sensitive to both composition
of the grains and their sizes (and thus radiative efficiency).  And the exact
cutoff in the spectrum depends sensitively on grain size in another way: aside
from the effect on $T$, small grains are not only hotter in a given radiation
field, they also deviate from blackbody emission at given $T$ as well.

So why does the $2\mu$ emission radius follow $L_{\rm BBB}$ so perfectly? I think it's
easy to understand why the grain properties are very generic. Far from the inner
edge of the torus, there is a great
variety of grain size and composition, and each grain has its own Rsub. But the
torus is a winnowing machine: as the grains work their way inward, only those
with the largest Rsub survive. The inner region must be made of the largest,
most refractory grains, and these may well be the same in all objects. The
closest grains are the toughest and they are just those large enough to emit
with near blackbody efficiency.

I think this very likely explains the special AGN extinction curve deduced from
the exceptionally well-designed study of Gaskell \etal\ (2004) \cite{Gaskell:2004}; that study
compared the spectra of radio quasars from complete samples of isotropically
(radio lobe) selected objects, as a function of core dominance. Arguably, given
the VLBI properties, it matches objects subject only to the validity of special
relativity and the Copernican principle. In fact, very remarkably, it's shown in
that paper that the spectral shape differences are truly due to reddening
because one can make curves based on comparing 6 emission lines from Ly $\alpha$
to Ha, which match the curves from the continua in both shape and normalization!

Now the fun really begins. None of the reverberation lag papers actually plots
lag vs.\ BBB luminosity! Instead, for the latter, a convenient proxy is used:
the (rest) $V$ magnitude! {\it This is not reasonable in principle} and the tiny dispersion
in the relation must mean that the bolometric correction from $M(V)$ to $L_{\rm BBB}$
must itself have a very small dispersion!

The $V$ band indicates the flux at the very bottom of the BBB, very far from the
peak in $\nu L_\nu$.  
If the BBB arises as optically thick thermal radiation from a fixed area, then
the characteristic temperature depends sensitively on the black hole mass and
Eddington ratio of luminosity. That is, $T \propto L^{1/4} M^{-1/2}$.
The bolometric correction from $V$ to BBB then depends very sensitively on the
characteristic temperature.

Thus for models of the type described, a very small dispersion would require
an unrealistically small range in $L^{1/4}M^{-1/2}$ at any given $M_V$.
I haven't proven
quantitatively here that this leads to a contradiction, but instead leave
that as an exercise.

\section{Part 7. Cygnus A Keeps on Giving --- recent results on our glorious prototype}

We are so fortunate to have such a prize as Cygnus A! Without checking the
folklore again, I'll tell you it's very likely at least roughly true that :

\textbullet\ \ Despite the volume element and strong cosmological evolution, you'd need to go
to a redshift of 1 to find another of it's radio luminosity. The redshift of
Cygnus is only 0.056.  It's claimed that in only one of 10,000 realizations of
our universe would we be so lucky, and we think it's largely a result of its
rich environment (Barthel and Arnaud) \cite{Barthel:1996}.

\textbullet\ \ A slightly inebriated expert assured me that this legend is true: 
if Cygnus is above the horizon, you can map it in a VLA sidelobe, independent
of the pointing. Even before the upgrade!

In this little section I'll recount some results from our group over the past
several years. First, our leader at the late Sofia airborne observatory, Enrique
Lopez-Rodriguez, has helped build the only sensitive far-IR polarimeter this
planet has ever deployed, then heroically and repeatedly soared into the
stratosphere to make various discoveries, not least that the far-IR emission
of Cygnus is 10\% polarized. This is an unprecedented result in the field, and
the interpretation came as a surprise to many.

Certainly one's first thought might be synchrotron emission, and the IR SED had
been modeled as such in multiple papers. I wasn't convinced by the models
because they were physically possible but not physically plausible. They
required extreme and fortuitous homogeneity and global parameters to fit the
very sharp sub-mm rise and the decline from the mid-IR to the near-IR. 
With the far-IR polarization measurement, it became clear that this is really
emission from dust grains aligned by a globally coherent magnetic field. In
the submillimeter region, the polarized flux of Cygnus A rises extremely
steeply with frequency like the total sub-mm flux, the signature of dust
{\it emission}. (The
near-IR polarization is similar in magnitude but somewhat dissimilar in
position angle, in fact falling {\it exactly} perpendicular to the radio jet.)

It seems very likely, especially considering analogous behavior in other
objects, that the near-IR polarization is none other than the result of
polar scattering of the unified model. That's true of the optical too, where
Ogle \etal\ \cite{Ogle:1997} have revealed enormously broad lines in polarized flux. The optical
scattering is only seen on scales of several arcsec.

While that large-scale scattering component isn't seen easily in the near-IR,
the polar (perpendicular) scattering is detectable with very high surface
brightness on sub-arcsec scales at $2\mu$ because this light penetrates a big
kpc-scale dust lane, very similar to the situation in Centaurus A. They are
discussed together with a third very similar case, 3C223.1, in Antonucci and
Barvainis 1990 \cite{Antonucci:1990}.


We (Enrique) have gone on to observe a heterogeneous handful of AGN, both
radio-quiet and radio-loud, finding so far a perfect correspondence between
those types and the detection and magnitude of polarization of the far dust
emission polarization \cite{Lopez-Rodriguez:2023}. This remarkable result must be explored further to
characterize it more fully and accumulate definitive statistics and, despite
the loss of Sofia, prospects for that are bright. The reason is that mighty
ALMA can cover enough of the crucial sub-mm range in many objects to look
for the extremely rapid rise in polarized flux that proves the origin as dust
emission.

This powerful discovery, if verified fully, is arguably our major clue to the
reason for the radio-loud vs quiet classes. Our theoretical colleagues were
sufficiently motivated to explore interpretations in our second paper. In
particular it's logically possible that the globally magnetized accretion flow
on pc scales ultimately manifests in a relativistic or sometimes slow
Seyfert-like jet; or that the opposite is true, and the jet instead
magnetizes the dusty gas. Much exciting work to do for all here.

Next, we are close to settling the nature of the gas in the $N(H)\sim3\times10^{23}$~cm$^{-2}$
X-ray column, a typical value for Type 2 objects. Of course the X-rays really
measure the heavy elements primarily, and much of that column resides in the
dust grains themselves. But is the accompanying gas in the atomic or molecular phase?   

Struve and Conway \cite{Struve:2010} have published beautiful and fascinating results from
H I 21~cm observations with VLBI.
Even the innermost {\it counter-jet} continuum component just $45$~milliarcsec
or $\sim50$~pc from the core
shows very strong H I absorption, with an equivalent width sufficient to
account for such a column density if one assumed a reasonable spin
temperature of $1\times10^4$~K. Yet the absorption in front of the radio core
is nearly two orders of magnitude smaller, and unable to account for the X-ray
columns unless T-spin is $1\times10^6$~K, which would be impossible to explain
in our view. Yet that's what the X-ray column measures for the core itself!

For many reasons, most importantly that such high columns are only observed in
Type 2 AGN, it's thought that the high X-ray columns are associated with the
torus-like obscurer of the U M rather than say atomic gas inside the
sublimation radius. Thus, absent sufficient atomic hydrogen column, we turn to
the possibility of a molecular torus as the default possiblity.
This must involve some exotic conditions however, because low-J CO rotational
absorption lines have long been sought without success against the core in the
millimeter region. The most recent limits come from ALMA observations at the
position of the $J = 1$ to 2 transition at rest-frame frequency of 230~GHz
(Carilli \etal\ 2022 \cite{Carilli:2022}, and pc). No absorption is seen,
and if the excitation of the low J states were thermal it'd be very conspicuous.   


It's been suggested based on much poorer limits that the low J states are
depopulated as we are driven to extremely high rotation temperatures by the
radiation of the radio core itself (Maloney \etal\ 1994) \cite{Maloney:1994}.
Great compactness is required, with almost the entire column within a few pc
of the very compact 1.5~mm core. For comparison, this sublimation radius for
blackbody grains which evaporate at 1500~K is estimated at $0.14$~pc.
Once again we think that ALMA can test this with a search for absorption in
a CO line near the atmospheric opacity cutoff. At that frequency the
solid-angle averaged brightness temperature is much reduced relative to that at
230~GHz.


Finally Carilli \etal\ \cite{Carilli:2022} report a beautiful-looking candidate torus in radio
continuum emission. The proposed radiation mechanism of Bremsstrahlumg doesn't
seem tenable (Bagul \etal\ 2023), the nature of the detected torus is still in
need of elucidation.

\section{Summary}
That was an arduous trek but we'll take the bus back.

Part 1. Spectroscopic Unification

The polarization properties of narrow line (Type 2) radio galaxies.  Torus and mirror. 

This Spectroscopic Unification seems to apply to most or all Seyfert galaxies
and quasars (including radio loud ones). But at low to medium radio luminosity,
there are plenty of radio galaxies which lack signs of copious realtime accretion,
namely the Big Blue Bump continuum and the Broad Emission Lines.

Part 2.  Beaming Unification by Orientation in the Radio Loud Objects 

Many properties of radio loud AGN, especially those of the Blazar class, are
very nicely explained with an orientation-based Unified Model and bulk
relativitic motion. Those with the directly visible or hidden BBB/BLR comprise
one group, and those without comprise another. The many nuances are described
systematically in e.g. Antonucci 2012.

Part 3. The Peculiar AGN Reddening Law

Direct arguments are given that lobe-dominant radio quasars and radio
galaxies must be intrinsically the same as matched blazars, only differing
in orientation. We select a complete sample based on ($\sim$isotropic) lobe flux,
and then we can unify the sample objects of the various subtypes with high
confidence. 

Dividing the composites as described produces an anisotropy due to
extinction which is highly chromatic in the visible, but quite flat in the UV.
This indicates a dearth of small grains, as expected qualitatively from the
winnowing action of the torus.

This reddening curve is then confirmed in a spectacular and independent way:
we showed that using the six emission lines from Ly $\alpha$ to H$\alpha$, we
make a new curve. It is identical in shape and normalization to the continuum
curve!

None of this precludes additional reddening by more normal dust in the host,
and we find much evidence for it, especially in radio-quiet Seyfert galaxies,
in keeping with their spiral hosts.

Part 4. The Accretion Disk Model for the Big Blue Bump:  Epic Magical Thinking.

The standard thin accretion disk model was proposed I'm the late 1970s and 80s
to account for the Big Blue Bump continuum component, which dominates the
luminosity of AGN.

Cognitive dissonance was part of this model since it was
proposed. For example, the quasistatic assumption is comically at variance with
observation. Similarly a scattering atmosphere of a thin disk with internal
dissipation produces a polarization perpendicular to the radio jet, opposite
to what we see.  It was soon shown that actual data show no sign whatever of
the basic expectation that the maximum disk temperature is proportional to the
one fourth power of the luminosity and the negative square root of the mass.
Other powerful arguments followed, but the non-viable disk model bandwagon
meant nobody noticed! 

Part 5. The spectrum of the central engine, with atomic and dust emission
removed!

{\it Se Raser la Barbe!}

There is little correlation between my own assessment of my papers and the
attention they receive. This is the most remarkable example: the BBB is energetically
dominant, but over most of its wavelength range from the near-IR to the EUV, it's
contaminated by reprocessed emission. We removed it over much of that range, and
the message can be extended. But the work received virtually no notice. M.~Kishimoto
deserves the lion's share of the credit. 

Luckily the polarization in most quasars is wavelength-independent in the
continuum, and we identified some cases in which the reprocessing features
are unpolarized.  
Therefore the polarized flux looks just like a noisy version of the total flux,
but without the contamination which heretofore prevented us from observing the
bare central engine spectrum. The results were rather amazing and our five
clear examples all behaved qualitatively the same way! They all show a broad
absorption feature starting at exactly the right location for the Balmer jump.
That strongly implies a contribution from an optically thick emitter with a
source function gradient like those in stars, indicating dissipation at large
optical depth.  

We then looked in the near-IR, to try to find out what the BBB is doing under
the ubiquitous hot dust emission.  In all five cases, the polarized flux turns
downward to reach $\nu^{+0.35\pm0.10}$ by $1.6\mu$ in the rest frame. The only
non-contrived explanation is that the dust is unpolarized as perhaps expected
for Type 1 (at low inclination), and the BBB turns over at those wavelengths.
That is another intrinsic spectral feature of the BBB (Kishimoto \etal\ 2008).

Recall though that microlensing observations indicate a surface brightness an
order of magnitude lower than required.  If these observations, at much shorter
rest wavelengths than the Balmer jump, are also from optically thick thermal
radiation, it must be that the effective areal covering factor is much less than one. 

Part 6.  Negating a mighty river of theory with a figure suitable for a T shirt.

The relationship between hot dust lags relative to BBB fluctuations has been
demonstrated to have {\it very} small dispersion. It follows the expected
relationship of lag time being proportional to the square root of luminosity,
but more important here is the remarkably small dispersion. 

{\it However}---$L_{BBB}$ can't be measured---so the observers used $M_V$ as a proxy!
{\it It worked extremely well.} But that's completely unexpected for a huge
superset of the standard disk models. It precludes any scenario in which the
characteristic temperature is set by $L^{1/4}M^{-1/2}$. In fact the
``temperature" can't depend {\it much} on anything at all, in most cases!

I consider that to be a profound result, but I leave the execution as an
exercise for the reader.

Part 7. The Magnificent Radio Galaxy Cygnus A---recent developments.

We've learned many new things about Cygnus A beyond its glorious and extremely
energy-rich radio lobes.

This prototype source fulfills all aspect of unified models, with high
perpendicular polarization of scattered light; a Type 1 spectrum in polarized
flux; beautiful bicones of polarized UV flux from HST; highly polarized dust
{\it emission} indicating a global magnetic field.

Other recent gossip is reported.

\section*{Acknowledgments}
Many astronomers provided comments on earlier versions of this paper. These
people suffered the most: R.~Barvainis, P.~Ogle, E.~Lopez-Rodriguez,
A.~Bagul, M.~Gaskell. A.~Bagul also kindly prepared most of the figures.
D.L.~Ceder made it much more intelligible.

\end{document}